\documentclass[twocolumn,letter]{jpsj3}
\title{Magnetic Ordering in V-Layers of the Superconducting System of Sr$_2$VFeAsO$_3$}

\author{Shunichi Tatematsu,$^1$ Erika Satomi,$^1$ Yoshiaki Kobayashi,$^{1,3}$\thanks{corresponding author(i45323a@cc.nagoya-u.ac.jp) } and Masatoshi Sato$^{1,2,3}$
}
\inst{$^1$Department of Physics, Division of Material Science, Nagoya University,
\\Furo-cho, Chikusa-ku, Nagoya 464-8602, Japan
\\$^2$Toyota Physical and Chemical Research Institute, Nagakute, Aichi 480-1192, Japan
\\$^3$JST, TRIP, Nagoya University, Furo-cho, Chikusa-ku, Nagoya 464-8602, Japan

}
\abst{Results of transport, magnetic, thermal, and $^{75}$As-NMR measurements are presented for superconducting Sr$_2$VFeAsO$_3$ with an alternating stack of FeAs and perovskite-like block layers. Although apparent anomalies in magnetic and thermal properties have been observed at $\sim $150 K, no anomaly in transport behaviors has been observed at around the same temperature. These results indicate that V ions in the Sr$_2$VO$_3$-block layers have localized magnetic moments and that V-electrons do not contribute to the Fermi surface. The electronic characteristics of  Sr$_2$VFeAsO$_3$ are considered to be common to those of other superconducting systems with Fe-pnictogen layers.
}

\kword{Fe pnictites, Sr$_2$VO$_3$, superconductivity, Hall coefficient, Seebeck coefficient, specific heat, $^{75}$As-NMR
}

\begin{document}
\maketitle

A new superconducting Fe pnictide, Sr$_2$VFeAsO$_3$, is composed of alternating stacks of FeAs and Sr$_2$VO$_3$ perovskite-like block layers. The superconducting transition temperature $T_c$ is $\sim $37 K.$^{1)}$ For this system, several band calculations have pointed out that both FeAs and V layers seem to have conduction electrons, and that the Fermi surfaces (FSs) do not have a significant nesting feature.$^{2-4)}$ This implies that Sr$_2$VFeAsO$_3$ may give us an opportunity to examine whether Fermi-surface nesting is important for the occurrence of superconductivity of Fe pnictides. On this point, Mazin has reported that, because the FSs constructed by only the Fe orbitals of Sr$_2$VFeAsO$_3$ are similar to those expected in other Fe pnictides, the nesting condition is also satisfied in the present system.$^{5)}$ However, it seems important to experimentally ensure whether or not the electrons of V ions are itinerant and really contributing to FSs, before studying the relation between Fermi-surface nesting and superconductivity.

We have carried out transport, magnetic, thermal, and $^{75}$As-NMR measurements, and found that although apparent anomalies in the temperature ($T$) dependences of the magnetic and thermal properties of Sr$_2$VFeAsO$_3$ exist at $\sim $150 K, no anomalies in the transport properties have been observed. On the basis of these results, we argue the electronic state of Sr$_2$VFeAsO$_3$ and answer the question regarding the contribution of V ions to FSs.

Polycrystalline samples of Sr$_2$VFeAsO$_3$ were prepared as described in refs. 1 and 6. SrAs powder was first obtained by annealing mixtures of Sr and As in an evacuated quartz tube at 850 $^\circ $C. Mixtures with proper ratios of SrAs, FeAs, SrO, Fe, and V$_2$O$_3$ were pressed into pellets, sealed in an evacuated quartz tube with Ti powder (Ti/Fe = 1/4), which probably acts as the reducing agent, and then fired for 10 h at 900 $^\circ $C and for 30 h 1050 $^\circ $C, successively.

\begin{figure}[tbp]
\begin{center}
\includegraphics[clip,width=9.0cm]{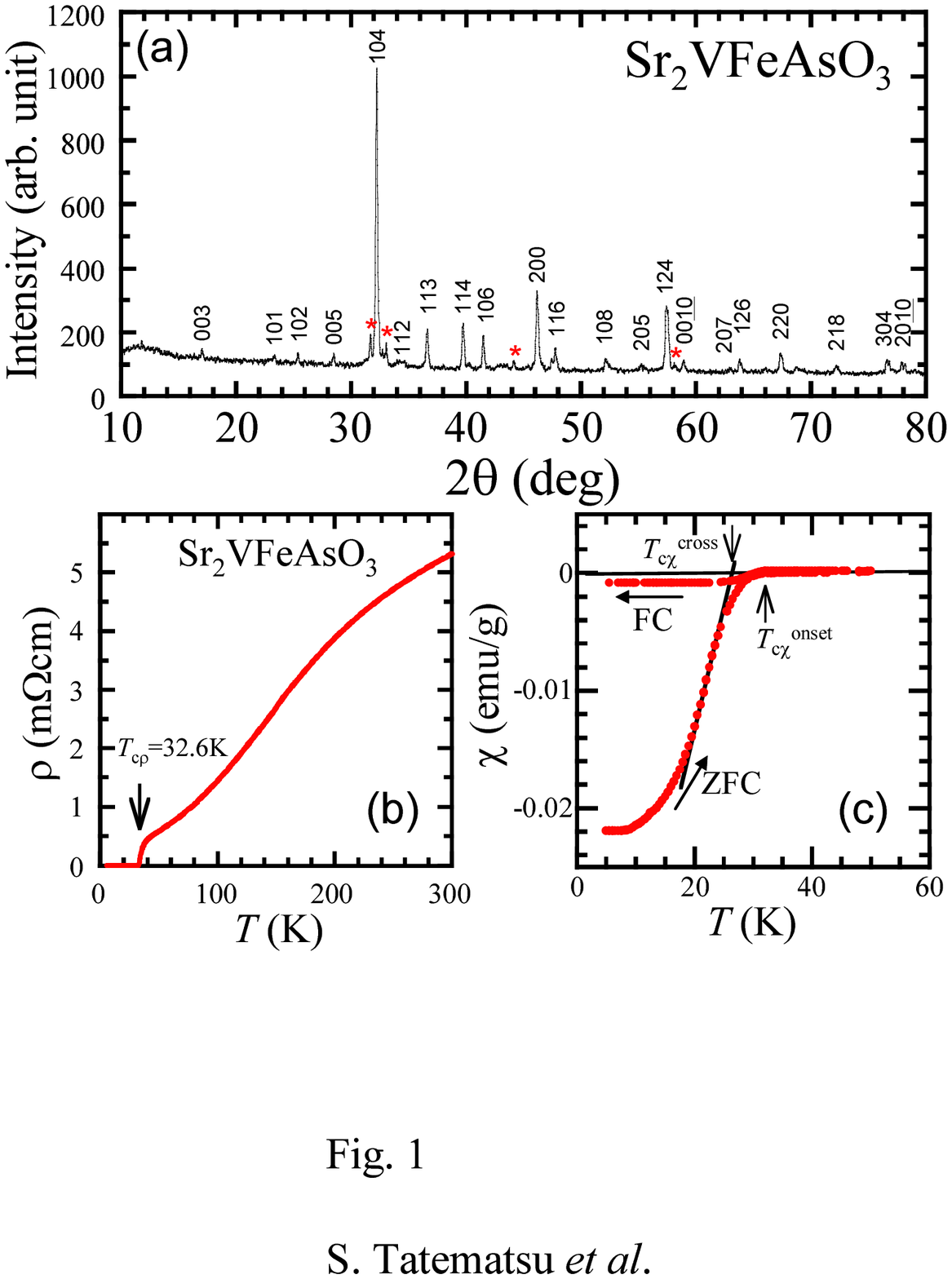}
\end{center}
\caption{(Color online) (a) Powder X-ray diffraction pattern of Sr$_2$VFeAsO$_3$, where the reflection indices are attached to the corresponding peaks of Sr$_2$VFeAsO$_3$. Minor impurity phases of Sr$_2$VO$_4$ or Sr$_3$V$_2$O$_7$ are shown by asterisks. (b) The electrical resistivity $\rho $ of a Sr$_2$VFeAsO$_3$ sample is shown against temperature $T$. $T_c$ estimated by the method described in ref. 11 is 32.6 K. (c) The superconducting diamagnetism at 10 G is shown for ZFC and FC in the form of an $(M/H\!\!-\!T)$ plot. $T_{c\chi} ^{cross}$ is estimated as shown in the figure.}
\label{fig1}

\end{figure}

The X-ray powder diffraction pattern of one of the obtained pellets with CuK$\alpha $ radiation at a step of 0.01$^\circ $ of the scattering angle 2$\theta $ is shown in Fig.\ref{fig1}
(a), where the reflection indices are attached to the corresponding peaks of Sr$_2$VFeAsO$_3$, and the asterisks indicate the peaks from impurity phases of Sr$_2$VO$_4$ or Sr$_3$V$_2$O$_7$.$^{7, 8)}$ The lattice parameters $a$ and $c$ were estimated to be 3.9329(1) and 15.6703(18) \AA, respectively.

The magnetic moments $M$ were measured at magnetic fields $H$ of 10 Oe and 1 T using a Quantum Design MPMS. The electrical resistivity $\rho $ was measured by the four-terminal method with increasing temperature $T$ at $H$ = 0. The thermoelectric power $S$ was measured by a dc method, where the typical temperature range between two ends of the sample was 0.2$-$2 K, depending on the temperature region. Details of the measurements are described in refs. 9 and 10. The specific heat $C$ and the Hall coefficient $R{\rm_H}$ were measured from 5 to 300 K using a Quantum Design PPMS. In the measurements of $R{\rm_H}$, $T$ was increased stepwise under $H$ = 7 T, where the sample plates were rotated around the axis perpendicular to the field. $^{75}$As-NMR measurements were carried out using the standard coherent pulse method at a magnetic field $H$ of $\sim $6 T using a fix frequency $f $ = 53.53 MHz.

The resistivity $\rho $ is plotted against $T$ in Fig. \ref{fig1}(b). From the data, the superconducting transition temperature $T_{c\rho }$ is determined to be 32.6 K by a method described in ref. 11 in detail. The superconducting diamagnetism was measured at $H$ = 10 Oe under both conditions of zero field cooling (ZFC) and field cooling (FC), and the data are shown in Fig. 1(c) in the form of a $\chi\text{ }(= M/H)\!-\!T$ plot. The temperature of superconducting transition, $T_{c\chi }^{cross}$, defined by the extrapolation shown in Fig. \ref{fig1}(c) is 26.6 K, while the onset value  $T_{c\chi }^{onset}$ obtained from the $\chi\!\!-\!T$ curve is 32.5 K, which is almost equal to  $T_{c\rho }$.

\begin{figure}[tbp]
\begin{center}
\includegraphics[clip,width=7.0cm]{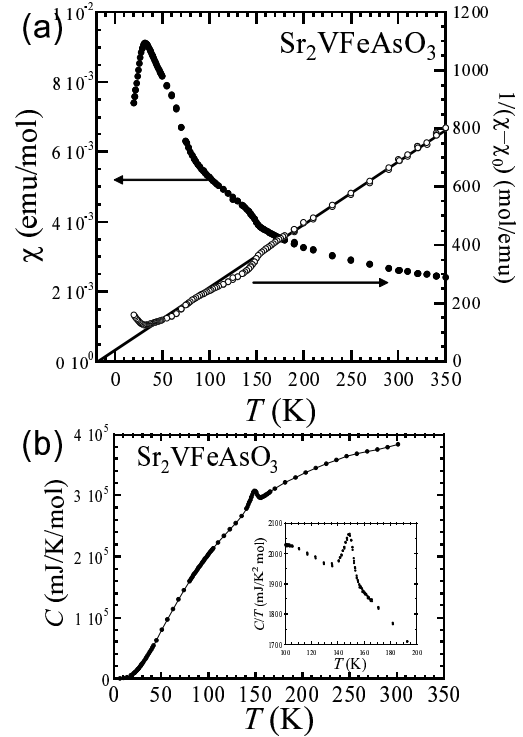}
\end{center}
\caption{(a) The magnetic susceptibility $\chi $ defined by $\chi  =M/H\text{ }(H =1\text{ }T)$ is plotted against $T$ by solid circles. Open circles show the plot of $1/(\chi -\chi _0)$ against $T$, where $\chi _0$ is the $T$-independent part of $\chi $. The straight line shows the result of Curie-Weiss fitting. (b) Specific heat data $C$ are plotted against $T$. In the inset, the $C/T\!-\!T$ curve is shown with magnified scales around 150 K.}
\label{fig2}

\end{figure}

The $T$ dependence of $\chi\text{ }(= M/H)\!-\!T$ measured at $H$ = 1 T under ZFC condition is shown in Fig. $\ref{fig2}(a)$. It obeys the Curie-Weiss law above $\sim $150 K as shown by the linear $T$ dependence of $1/(\chi -\chi _0)$ ($\chi _0$ is a $T$-independent term), where the effective magnetic moment 
$ \mu _{\rm eff}$
 is estimated to be 1.90 $\mu _{\rm B}$ and the Weiss temperature is -15 K.  $\mu _{\rm eff}$ is between 1.73  $\mu_{\rm B}$ expected for V$^{4+}$ (spin $S =1/2$) and 2.83 $\mu_{\rm B}$ expected for V$^{3+}$ ($S =1$). With decreasing $T$, the absolute slope $|{\rm d}\chi /{\rm d}T| $ shows an anomalous increase or deviates from the Curie-Weiss law at $\sim $150 K. Because the specific heat $C$ has an unambiguous peak at $\sim $150 K, as shown in Fig. 2(b), we can consider that this anomaly in $\chi $ is intrinsic. The impurity phase, Sr$_2$VO$_4$ or Sr$_3$V$_2$O$_7$, does not exhibit any anomaly in the $\chi\!-\!T$ curve at $\sim $150 K: For Sr$_2$VO$_4$, the Curie-Weiss $T$-dependence of $\chi $ is observed above 45 K and the $\chi $ anomaly is just found only at $\sim $45 K, while Sr$_3$V$_2$O$_7$ exhibits Pauli paramagnetic behavior.$^{7)} $ We will show later another evidence obtained by $^{75}$As-NMR indicating that the anomaly at $\sim $150 K is intrinsic. As shown in Fig. 2(a), the slope $|{\rm d}\chi /{\rm d}T| $ begins to show a slightly anomalous increase with decreasing $T$ below $\sim $90 K, this behavior may not be intrinsic.

The change in the entropy around $\sim $150 K is estimated using the anomaly in the 
$C/T\!\!-\!T$ curve [inset of Fig. 2(b)] to be $\sim\!\!\!1.6 \text{ }\rm{J/(mol\!\cdot\! K)}$, which corresponds to only 28\% (18\%) of $N_Ak_B\!\!\cdot\! ln2 \> (N_Ak_B\!\!\cdot\! ln3)$ for the magnetic ordering of localized spins of $ S = 1/2$ (or 1), where $N_A$ and $k_B$ are the Avogadro number and Boltzmann factor, respectively. Although this entropy change depends on the estimation of the phonon contribution to $C/T$, we think that it is smaller than the above value for $S = 1/2$ (or 1), because the rather sharp shape of $C/T\!\!-\!\!T$ does not bring about a very large error in the estimation. We will discuss the possible origin of this small value of the entropy change later. Because the $\chi\!\!-\!T$ curve obeys the Curie-Weiss law up to 800 K, there is no loss of entropy caused by the short-range order above $\sim 150$ K.

The Hall coefficient $R{\rm_H}$ and the Seebeck coefficient $S$ are plotted against $T$ in Figs. 3(a) and 3(b), respectively. These $T$ dependences are similar to those observed in the electron-doped system LaFeAsO$_{1-x}$F$_x$, indicating the effect of strong magnetic fluctuations.$^{12-15)}$ The peak value of $\mid \!\!S\!\!\mid $ is similar to that of a slightly overdoped sample of LaFeAsO$_{1-x}$F$_x$.$^{16)}$ No anomalies are observed in the $R{\rm_H}\!\!-\!\!T$ and $S\!\!-\!\!T$ curves at $\sim $150 K, where the magnetic susceptibility and specific heat exhibit apparent anomalies. Because $S$ is very sensitive to anomalous Fermi surface changes, we can say that there is no anomalous change in the Fermi surface of the present system at $\sim $150 K. This result does not depend on whether data are taken for poly- or single-crystalline samples, because, as is widely known, $S$ is hardly influenced by grain boundaries. From the observations described above, we can say the following. If the electrons of V ions contribute to the conductivity, the change of the electronic (magnetic) state in the V layers should be observed in the transport behaviors, at least in the thermoelectric power $S$. Therefore, the anomalies in $\chi $ and $C$ are considered to be confined to the V layers, and V ions do not contribute to the FSs.

\begin{figure}[tbp]
\begin{center}
\includegraphics[clip,width=8cm]{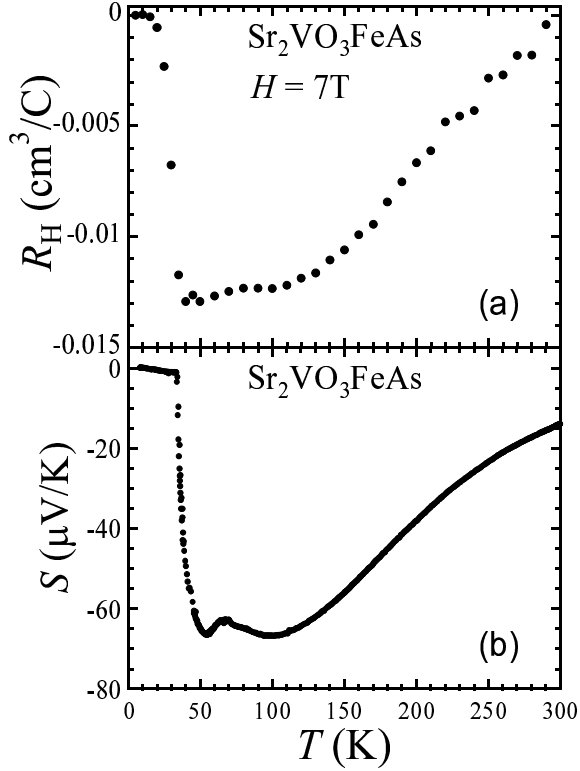}
\end{center}
\caption{Hall coefficient $R{\rm_H}$ and Seebeck coefficient $S$ are plotted against $T$ in (a) and (b), respectively.}
\label{fig3}
\end{figure}

To investigate the electronic state microscopically, we have carried out measurements of  $^{75}$As-NMR spectra and the $^{75}$As nuclear spin-lattice relaxation rate $1/T_1$. Figure 4(a) shows central transition lines ($I_z =1/2 \longleftrightarrow 1/2; I_z$ is the $z$ component of the nuclear spin) of the NMR spectra, for example, at four temperatures. With decreasing $T$ through $\sim $150 K, the shape in the low-$H$ part of the center line changes, and the center line shifts as a whole to the lower field side although the width of the center line does not change markedly. Since the width is explained by the second-order effects of the nuclear electric quadrupole interaction, the nuclear electric quadrupole resonance frequency $\nu_Q$ is roughly $T$-independent and the Knight shift increases with decreasing $ T$ through $\sim$150 K, indicating that the structural change around the As site does not take place at  $\sim$150 K, but that the magnetic environment around the As site changes. Consequently, we can exclude the structural transition at  $\sim$150 K. As shown in Fig. 4(a), the shape of the central transition line is almost unchanged in the $T$ region between  $\sim$150 and  $\sim$70 K and the line gradually broadens with decreasing $T$ below  $\sim$70 K. Because the magnetically ordered structure of the V site is not known, we do not know whether the hyperfine field at the As site from a few V sites should be canceled. However, the occurrence of an 
\begin{figure}[tbp]
\begin{center}
\includegraphics[clip,width=9.6cm]{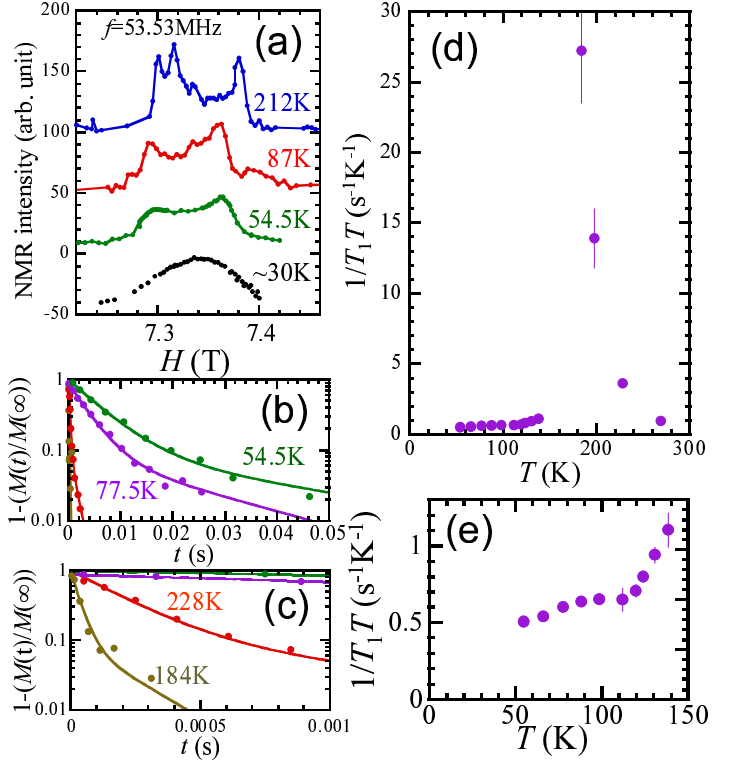}
\end{center}
\caption{(Color online) (a) Central transition lines $(I_z =1/2 \leftrightarrow 1/2)$ of the $^{75}$As-NMR spectra for the powder sample at four temperatures. (b) and (c) $^{75}$As-NMR relaxation curves are plotted against the time $t$ elapsed after the irradiation of the saturation RF pulses. The solid lines show the result of fittings of the theoretical curves to the corresponding data. (d) and (e) The $^{75}$As-$1/T_1T$ values obtained by the fittings are plotted against $T$.}
\label{fig4}
 \end{figure}
inhomogeneous internal field observed as the smearing of the centerline below  $\sim$70 K, 
may indicate that the magnetic structure is changing with  $T$. We presume that the change is understood by the spin-glass-like transition in a certain fraction of the system at  $\sim$150 K, almost simultaneously with the magnetic transition.
$^{75}$As-1/$T_1$ is measured at the position of the lowest-$H$ peak. In Figs. 4(b) and 4(c), the nuclear magnetization $M(t)$ is plotted against the time $t$ elapsed after irradiating saturation RF pulses. The $1/T_1$ values are estimated by fitting the theoretical curves, described for the nuclear spin as $ I=3/2$, $1-M(t)/M(\infty ) = 0.9\cdot  \mathrm{exp}(-6t/T_1) + 0.1\cdot \mathrm{exp}(-t/T_1)$, to the observed $M(t)$-data. The obtained $1/T_1T$-values are plotted against $T$ in Figs. 4(d) and 4(e), where we can clearly see that $1/T_1T$ exhibits a diverging tendency as the system approaches $\sim $150 K from above, and then decreases rapidly with decreasing $T$ in the $T$-region below $\sim$150 K. The $^{75}$As-nuclear spin-spin relaxation rate $1/T_2$ also shows similar behavior in the same $T$-region (not shown). 
These divergent behaviors of $1/T_1T$ and $1/T_2$ at $\sim $150 K indicate again that a certain magnetic transition takes place at $\sim $150 K. Then, together with the observations that $R{\rm_H}$, $S$, and $\rho $ exhibit no anomalies at $\sim $150 K, the present results indicate that the electrons of V ions do not contribute to the FSs of the system. 


Although the smallness of the change in the entropy might indicate that some of the electrons of the V-layer are itinerant, the absence of the anomalous transport behaviors at the transition point excludes such a possibility. Thus, some of the V sites undergo the spin-glass transition at $\sim $150 K.

As the superconductivity appears in the Sr$_2$VFeAsO$_3$ samples prepared by heating with a proper amount of Ti metal as a getter of oxygen, the carrier doping to the FeAs layers is considered to be caused by the oxygen deficiency, and randomness is introduced into the V-O layers; with that, a complex magnetic behavior may be observed. 
Sr$_2$CrFeAsO$_3$, which has a similar structure to Sr$_2$VFeAsO$_3$, has been reported to have similar magnetic features, that is, its Cr ions are in the Mott-insulating state, and have localized spins $(S =3/2)$, which order at 36 K.$^{17)}$ A similar situation is also observed in Sr$_2$VFeAsO$_3$.
Recently, the results of angle-resolved photoemission spectroscopy (ARPES) have been reported for the present system, where FSs are found to be similar to those of other Fe-pnictide superconductors.$^{18)}$ The present data obtained by the transport and NMR studies are consistent with the ARPES results. The recent band calculation carried out by taking account of the onsite Coulomb energy of the V atom also supports all the observations described here.$^{19)}$

In summary, we have measured the magnetic, transport, and thermal properties and also carried out $^{75}$As-NMR studies of Sr$_2$VFeAsO$_3$. The results of the studies show that V ions have localized moments and exhibit magnetic order at $\sim $150 K, although randomness effects introduced by oxygen deficiency seem to exist. The FSs of this system can be considered to have common characteristics to those of other Fe pnictides.

This work is supported by Grants-in-Aid for Scientific Research from the Japan Society for the Promotion of Science (JSPS) and Technology and JST, TRIP.

\end{document}